\documentclass[twocolumn,final]{elsarticle}
\usepackage{jasr}
\usepackage{framed,multirow}
\usepackage{array, booktabs, tabularx} 
\usepackage{amssymb}
\usepackage{latexsym}
\usepackage{amsmath}
\usepackage{caption}
\usepackage{xspace}
\usepackage{comment}
\usepackage[left]{}

\definecolor{newcolor}{rgb}{.8,.349,.1}

\usepackage{amsmath}
\usepackage{amssymb}
\usepackage{graphicx}
\usepackage{caption}
\usepackage{subcaption}
\usepackage{url}
\usepackage{xcolor}
\usepackage{multicol}
\usepackage[utf8]{inputenc}

\usepackage{placeins}
\usepackage[citebordercolor=white]{hyperref}
\biboptions{authoryear}

\journal{Advances in Space Research}
\begin{document}
\newcommand{\mirya}{Mirya-$\mu$1\xspace}

\verso{Da\u{g} \textit{et al.}}
\begin{frontmatter}

\title{The \mirya Cosmic Ray Detector: Features and First Year Observations}

\author[1,2]{Meryem K\"ubra \snm{DA\u{G}}\corref{cor1}}
\cortext[cor1]{Meryem K\"ubra DA\u{G}: 
  Tel.: +44 0191 334 3520;  
  meryem.k.dag@durham.ac.uk;}
\author[3,4,5]{Tolga \snm{G\"UVER}}
\author[6]{G\"uray \snm{G\"URKAN}}
\author[7]{Ecem \snm{AKTA\c{S}}}
\author[8]{Suat \snm{\"OZKORUCUKLU}}
\author[9]{Sindulfo \snm{AYUSO}}
\author[9]{Juan Jos{\'{e}}  \snm{BLANCO}}
\author[3,4]{Zahide Funda \snm{BOSTANCI}}
\author[2]{Beste \snm{BEG\.{I}\c{C}ARSLAN}}
\author[2,10]{Mustafa Turan \snm{SA\u{G}LAM}}
\author[11]{Ahmet \snm{POLATO\u{G}LU}}
\author[11,12]{Cahit \snm{YE\c{S}\.{I}LYAPRAK}}

%\address[2]{Affiliation 1, Address, City and Postal Code, Country}
\affiliation[1]{organization={Durham University},
                addressline={Centre for Advanced Instrumentation
Department of Physics},
                city={Durham},
                postcode={DH1 3LE},
                country={UK}}
\affiliation[2]{organization={Istanbul University},
                addressline={Institute of Graduate Studies in Science, Department of Astronomy and Astrophysics},
                city={Istanbul},
                postcode={34134},
                country={Turkiye}}   
\affiliation[3]{organization={Istanbul University},
                addressline={Observatory Research and Application Center},
                city={Istanbul},
                postcode={34119},
                country={Turkiye}}
\affiliation[4]{organization={Istanbul University},
                addressline={Faculty of Science, Department of Astronomy and Space Sciences},
                city={Istanbul},
                postcode={34119},
                country={Turkiye}}
\affiliation[5]{{Georgia Institute of Technology}
addressline={School of Physics, Georgia Institute of Technology, 837 State St NW},
                city={Atlanta},
                postcode={30332},
                country={Georgia, USA}}
\affiliation[8]{organization={Istanbul University},
                addressline={Faculty of Science, Department of Physics},
                city={Istanbul},
                postcode={34134},
                country={Turkiye}}
\affiliation[7]{organization={Istanbul University},
                addressline={Institute of Graduate Studies in Science, Department of Physics},
                city={Istanbul},
                postcode={34134},
                country={Turkiye}}   
        \affiliation[6]{organization={Istanbul University},
                addressline={Faculty of Science, Department of Physics, Nuclear Physics Division},
                city={Istanbul},
                postcode={34134},
                country={Turkiye}}

\affiliation[9]{organization={Department of Physics and Mathematics, Space Research Group, Universidad de Alcalá, Ctra. Madrid-Barcelona km 33,6},
                city={Alcalá de Henares},
                postcode={28871},
                country={Spain}}
\affiliation[10]{organization={Erciyes University},
                addressline={Science Faculty, Department of Astronomy and Space Sciences},
                city={Kayseri},
                postcode={38030},
                country={Turkiye}}          
\affiliation[11]{organization={Ataturk University},
                addressline={Science Faculty, Astronomy and Space Sciences Department},
                city={Erzurum},
                postcode={25050},
                country={Turkiye}}
\affiliation[12]{organization={Turkiye National Observatories},
                addressline={DAG},
                city={Erzurum},
                postcode={25050},
                country={Turkiye}}
\received{}
\finalform{}
\accepted{}
\availableonline{}
\communicated{}

\begin{abstract}
%%%
We introduce the \mirya Cosmic Ray Detector, the largest and only cosmic ray detector in T\"urkiye designed for space weather research. \mirya, modeled and built after the Muon Impact Tracer and Observer (MITO) \citep{MITO}, is located at the Eastern Anatolia Observatory (DAG) site of the Türkiye National Observatories in Erzurum, Türkiye, at an altitude of 3099 meters. This elevation positions \mirya among the highest-altitude cosmic ray detectors globally.
The detector consists of two stacked scintillator counters, each measuring 1$\times$1 meters, separated by a vertical distance of 1.36 meters. Each scintillator is monitored by four H1411 Hamamatsu photomultiplier tubes, enabling precise detection and measurement of light by incident cosmic rays. In this study, we present the data collected throughout 2024, which includes the detection of two Forbush decrease events in March and May 2024. These significant detections demonstrate the capability of \mirya to contribute valuable data for space weather research, establishing its potential as a critical instrument for cosmic ray studies in the region.\\
%%%%
\end{abstract}

\begin{keyword}
%% Keywords
\KWD Cosmic Rays\sep Muon Detector\sep Space Weather
\end{keyword}

\end{frontmatter}

\section{Introduction}
\label{sec1}

Cosmic rays are particles from deep and nearby space that are continuously entering our solar system and interacting with Earth's atmosphere. The energy range of cosmic rays varies between 10$^{9}$~eV and 10$^{21}$~eV, making them one of the highest-energy phenomena in the Universe. Due to their charged nature, they encounter magnetic fields during their journey and deviate from their original trajectories \citep{Stanev2010, UHECRspektrum5}. Cosmic rays with energies above $10^{15} \, \text{eV}$ are generally thought to be accelerated by high-energy events occurring outside our galaxy, while lower-energy cosmic rays originating from sources within our galaxy.  Before reaching Earth, these particles are referred to as primary particles. Approximately 90\% of the primary particles are protons, 9\% are helium nuclei, and 1\% consist of electrons and other heavy atomic nuclei. When primary particles collide with heavy atomic nuclei in the upper atmosphere, they decay into secondary particles such as pions ($\pi^\pm$), kaons ($K^\pm$), and muons ($\mu^\pm$) \citep{GRIEDER20011}. Muons ($\mu^-$ and $\mu^+$) can be easily detected by  using scintillators and are the most numerous charged particles at sea level.

Cosmic rays have recently gained significant importance in space weather monitoring. During their journey through the heliosphere, cosmic rays are modulated by solar activity, including events such as coronal mass ejections (CMEs) and shock waves \citep{Cane_CME}. As a result of this modulation, cosmic rays are considered crucial for space weather studies in several respects. Firstly, cosmic rays possess long mean free paths, which are sufficient to carry precursor signatures of disturbances occurring between the Sun and Earth. Furthermore, their Larmor radii are larger than Earth's magnetosphere but smaller than the radius of typical disturbances in the heliosphere. Consequently, cosmic rays can travel without being influenced by Earth's magnetosphere while still carrying information about solar activity. Prior to geomagnetic events, the isotropy of muon flux is often disturbed, indicating incoming geomagnetic disturbances \citep{MITO, kuwabara2006}.

Forbush decreases (FDs) are transient and sudden reductions in cosmic ray intensity measured on ground-based detectors. This phenomenon is generally associated with the Sun's magnetic activity, including coronal mass ejections (CMEs), magnetic clouds, and interplanetary shock waves \citep{Cane2000, Fushishita2008}. It is believed that the intense magnetic fields carried by such solar activity interact with galactic cosmic rays, leading to their scattering or deflection, ultimately causing a reduction in the flux reaching Earth. This makes FDs an important event proposed for space weather monitoring \citep{Cane2000, Stanev2010}. However, while the connection between CMEs and FDs is well documented, the precise mechanisms governing the onset and magnitude of these decreases require further observational and theoretical investigation \citep{refId0, 2refd0}.

The detection of cosmic rays is typically carried out using neutron monitors or muon detectors. The Global Muon Detector Network (GMDN), consisting of muon detectors distributed across the globe, and the Neutron Monitor Database (NMDB), consisting of neutron monitors, play a critical role in sharing data between detectors located in different regions \citep{GMDB, nmdb}. These networks enable the observation of how solar activity affects cosmic ray flux globally. Additionally, muon detectors have been shown to provide earlier precursor detections compared to neutron monitors due to their sensitivity to a wider energy range (up to several hundred GeV). Moreover, muon detectors have been operational for a longer period than neutron monitors, providing more extended datasets for studying long-term trends in anisotropies and other cosmic ray variations as a function of solar cycles \citep{Duldig2000}.

We here introduce the \mirya Cosmic Ray Detector,  based on the design of \textbf{M}uon \textbf{I}mpact \textbf{T}racer and \textbf{O}bserver (MITO) cosmic ray detector, operated in Antarctica by the University of Alcal\'a, Spain \citep{MITO}. Like MITO, \mirya consists of two scintillators, each with an area of 1~m$^{2}$, and eight PMTs installed at the four sides of each scintillator. MITO was chosen as a reference due to its straightforward design, which requires fewer components and electronics compatible to the  other systems. Additionally, MITO can determine particle trajectories by comparing pulse heights in each scintillator, providing both cosmic ray count rates and impact position data \citep{MITO}.

\mirya is a key component of T\"urkiye's first space weather monitoring station. It aims to provide space weather predictions by investigating the relationship between solar activity and cosmic ray intensity/isotropy, such as Forbush decreases. The detector has been operating almost continuously for over a year at the T\"urkiye National Observatories' Eastern Anatolia Observatory (DAG) site in Erzurum, T\"urkiye. In Section \ref{sec:ins} we introduce the details of \mirya and its components. In Section \ref{sec:loc} we provide information on the observatory site where \mirya is located. Finally, we show the first year data in Section \ref{sec:fyear_obs} and conclude in Section \ref{sec:conc}.

\section{Instrument Design and Components}
\label{sec:ins}
As previously mentioned, the \mirya detector is built based on the MITO cosmic ray detector \citep{MITO}. The MITO detector was chosen as a model due to its minimal number of components and feasibility for construction under laboratory conditions. Additionally, this design is suitable and sufficient for establishing a system dedicated to space weather monitoring. Having similar systems around the world can also help overcome inter-calibration issues when the same events are observed. \autoref{tab:detector_comparison}  we give a comparison of \mirya and MITO in Antarctica and Tenerife.

\begin{table*}[h]
    \centering
    \caption{Comparison of Mirya-m1 and MITO Cosmic Ray Detectors in Antarctica and Tenerife.}
    \renewcommand{\arraystretch}{1.2} 
    \setlength{\tabcolsep}{6pt} 
    \begin{tabular}{lccc}
        \toprule
        \textbf{Parameters} & \textbf{\mirya} & \textbf{MITO Antarctica} & \textbf{MITO Tenerife} \\
        \midrule
        Altitude (m) & 3099 & sea level & 2364\\
        Average Count Rate (8PMT) & 2993 &  1751 & 1911\\
        Scintillator Size (m$^2$) & 1.0 & 1.0 & 1.0 \\
        Scintillator Separation (m) & 1.353 & 1.365 & 1.493\\
        Scintillator Model & EJ-200 & BC-400 & BC-400\\
        PMT Model & H11411 & R2154-02 & R2154-02\\
        Number of PMTs & 8 & 8 & 8 \\
        \bottomrule
    \end{tabular}
    \label{tab:detector_comparison}
\end{table*}

\mirya is fundamentally a scintillation detector comprising two EJ200 scintillators, each measuring 1~m $\times$ 1~m (100~cm $\times$ 100~cm $\times$ 5~cm, made of poly-vinyl-toluene with 64\% anthracene), and eight H11411 Hamamatsu photomultiplier tubes (PMTs). 

\mirya TOP and \mirya BOTTOM represent the two identical detection systems of the detector, similar to MITO-top and MITO-bottom. Both systems are horizontally stacked in a four-layer rack configuration. The middle two layers of the rack are filled with lead sheets to eliminate  
low-energy particle signals. The working mechanism of \mirya is no different from a typical scintillation detector. First, light generated by a charged particle passing through \mirya TOP exits through the four narrow side faces of the scintillator and is directed to four PMTs via air light guides. If the same particle has sufficient energy, it passes through the lead layer and produces scintillation in \mirya BOTTOM. Each PMT collects light from the four corners of each scintillator (TOP and BOTTOM) and generates a pulse. The pulses from each PMT output are sent to a C7319 Hamamatsu amplifier, where it is amplified and inverted as required, and then sent to the electronic readout circuit. The electronic circuit was also produced considering the MITO design principles \citep{MITO}.

The \mirya detector contains two highly sensitive and fragile scintillators, requiring complete isolation from any light other than scintillation. Therefore, a specially designed mechanical system was established. The two profile racks housing the \mirya TOP and BOTTOM scintillators were fully covered with black cardboard and high-voltage electrical tape \citep{dag2023mirya}. The inner compartment housing the scintillator was completely lined with white plexiglass. White color was used to achieve Lambertian reflection \citep{Lambertian} and thus to reflect scintillation light without loss. The joints where the plexiglass meets the profile were first covered with aluminum tape and then with electrical tape, ensuring that all areas inside were carefully sealed to prevent light ingress.
The four corners of the profiles supporting each scintillator system were left open to integrate the PMT and amplifier systems. In these openings, PMT holders, produced in the Istanbul University Observatory Laboratory and designed in the Department of Physics, were placed. The PMT holder designs were fabricated using 3D printers \citep{dag2023mirya}.

\begin{figure*}[t]
\centering
    \includegraphics[scale=0.09]{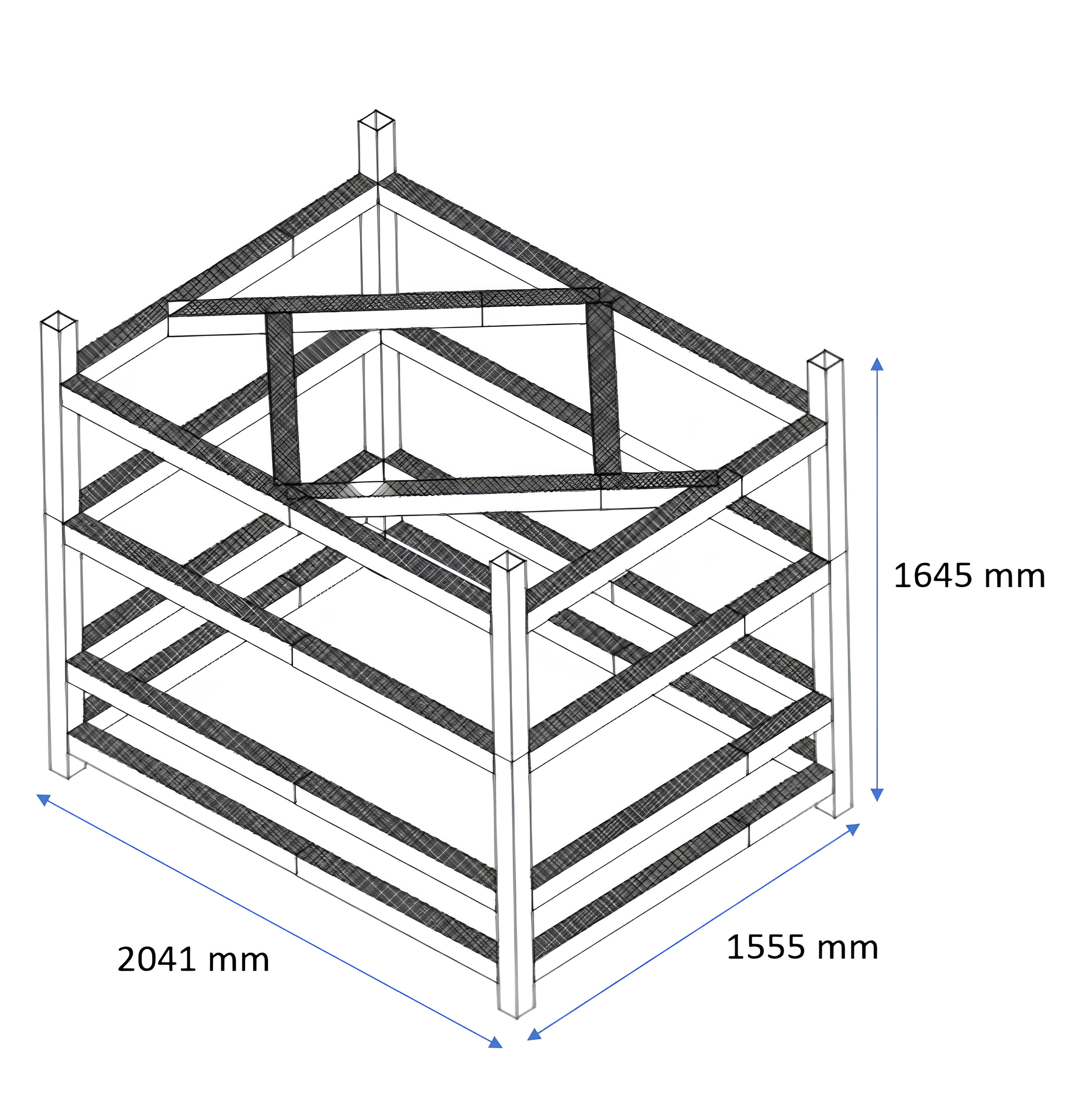}
    \includegraphics[scale=0.28]{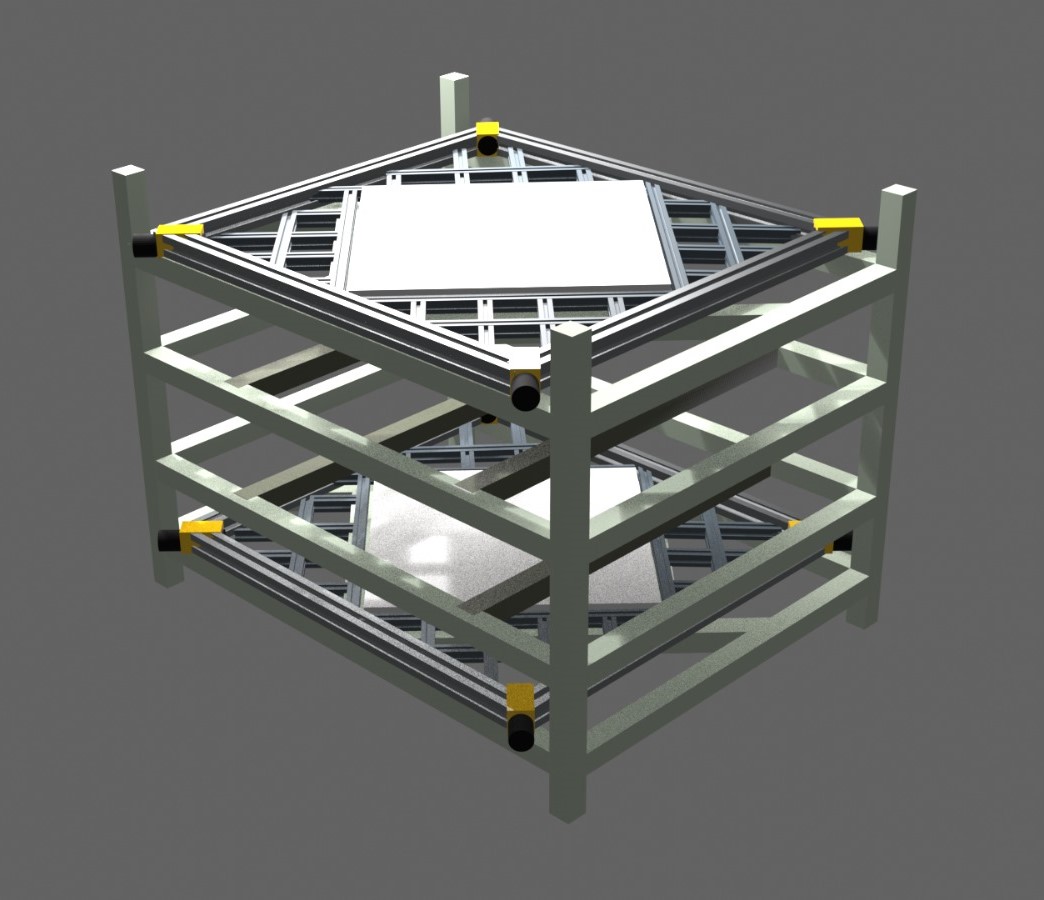}
    \caption{Isometric view of the four-layer shelves with dimensions labeled. In the left panel only the location of the scintillator on the upper layer is shown. Right panel shows the main detector structure as a 3D model. On the two layers in between the scintillators, we have lead layers, which are not shown for clarity. }
    \label{fig:structure}
\end{figure*}

The square iron profiles containing the two separate systems were also incorporated into a four-layer cast iron shelving system (\autoref{fig:structure}). The two layers in between the scintillators house lead with total 5.5 and 6.5~cm thicknesses. These are not individual solid lead plates but rather in total 40 layers of lead each having 2mm thickness. In the MITO design, a 10 cm thick lead layer was reported to provide performance equivalent to the neutron detector used in MITO \citep{MITO}. According to GEANT4 simulations conducted during the MITO setup phase, this lead layer acts as a filter, blocking the passage of charged particles above 200 MeV for muons, 160 MeV for protons, and more than 5 GeV for electrons \citep{MITO}. The \mirya TOP and \mirya BOTTOM systems are each wrapped by additional light-proof fabric on the shelves they are placed on \citep{dag2023mirya}. We note that a complete simulation of the \mirya detector response is beyond the scope of this paper and detailed GEANT4 simulations are currently underway. In particular, the GEANT4 simulations previously conducted for the MITO detector which shares an identical design with \mirya are largely applicable \citep{MITO}.

Similar to the \cite{MITO} design, the two layers and 8 PMTs provide a setup, which can be used to test the isotropy of the incoming particles. Additionally, the separation in between the scintillators allows for a constrained solid angle. Following the calculation in \cite{MITO}, we can calculate the solid angle subtended by \mirya as 1.445~sr, given the total separation of 1353~mm, very similar to MITO. Overall GEANT4 simulations show that the angular resolution reached by the system is $\pm3.9^{o}$ in the zenith angle and $\pm16.2^{o}$ in the azimuth angle for MITO \cite{MITO}, given the identical configuration \mirya has a similar angular resolution as well.

Because of this configuration, while operating the events that trigger all of the PMTs are not only going to be the most energetic ones, since they are able to pass through the lead layers, but will also be coming from the zenith of the detector with a solid angle of 1.445~sr. The events that only trigger the upper layer will contain both lower energy particles and particles that may be coming from a wider angle. We note that with such a configuration studies related to the cosmic ray isotropy can also be done with \mirya in the same way it is done with MITO.

\subsection{Data Acquisition}

The electronic readout circuit of \mirya is entirely based on the MITO design \citep{MITO} and was developed in collaboration with the University of Alcal\'a. It was manufactured at the DEKA PCB factory in Istanbul, T\"urkiye \footnote{\url{https://www.dekaelectronic.com/en/}}. All electronic module tests were conducted at the Istanbul University Observatory Laboratory, in coordination with the Physics Department and the MITO team of the University of Alcal\'a.

\begin{figure}
    \centering
    \includegraphics[width=\linewidth]{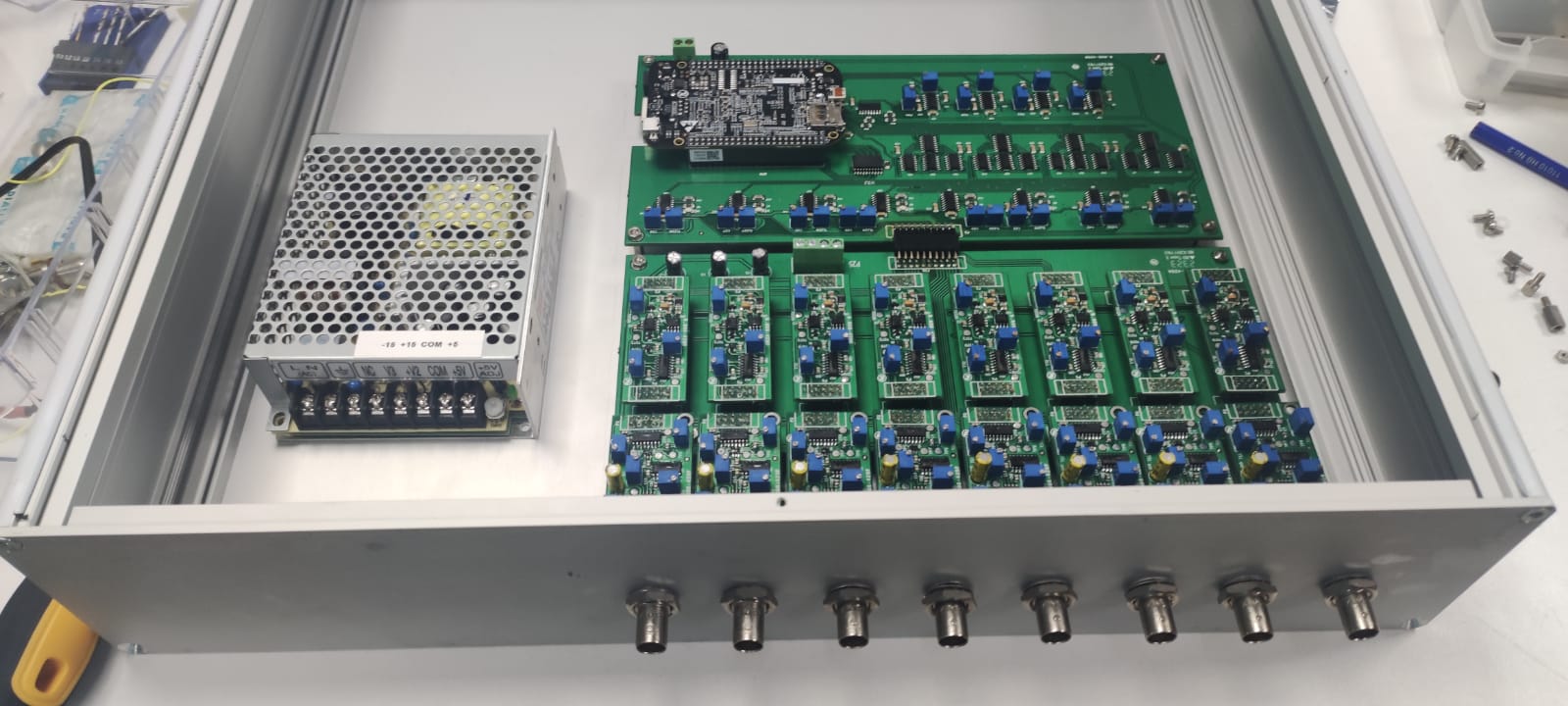}
    \captionof{figure}{\mirya electronic module in a protected box at Istanbul University Physics Laboratory,
2023 \citep{dag2023mirya}.} 
    \label{fig:daq} 
\end{figure}

The MITO data acquisition system (DAQ) consists of analog and digital processing modules, enabling parallel data analysis. However, \mirya adopted only the analog module, known as Software-coincidence Acquisition System (SAS), while the digital module, Adaptable and Reconfigurable Acquisition Concept for Nuclear Electronics (ARACNE) \citep{blanco2021orca, ayuso2016coincidence}, was excluded. The analog module (\autoref{fig:daq}) processes PMT signals by limiting the detectable particle energy range through lower-level (LLD) and upper-level (ULD) discriminators, effectively filtering out noise and high-energy particles \citep{MITO}.

In \mirya, as in MITO, a Beaglebone Black\footnote{\url{https://docs.beagleboard.org/boards/beaglebone/black/index.html}} single-board computer (SBC) manages data processing. The detected pulses from all eight channels are captured, analyzed for overlap and amplitude, and stored for further analysis \citep{MITO}. 

\subsection{Voltage Control}

\begin{figure}
    \centering
    \includegraphics[width=\linewidth]{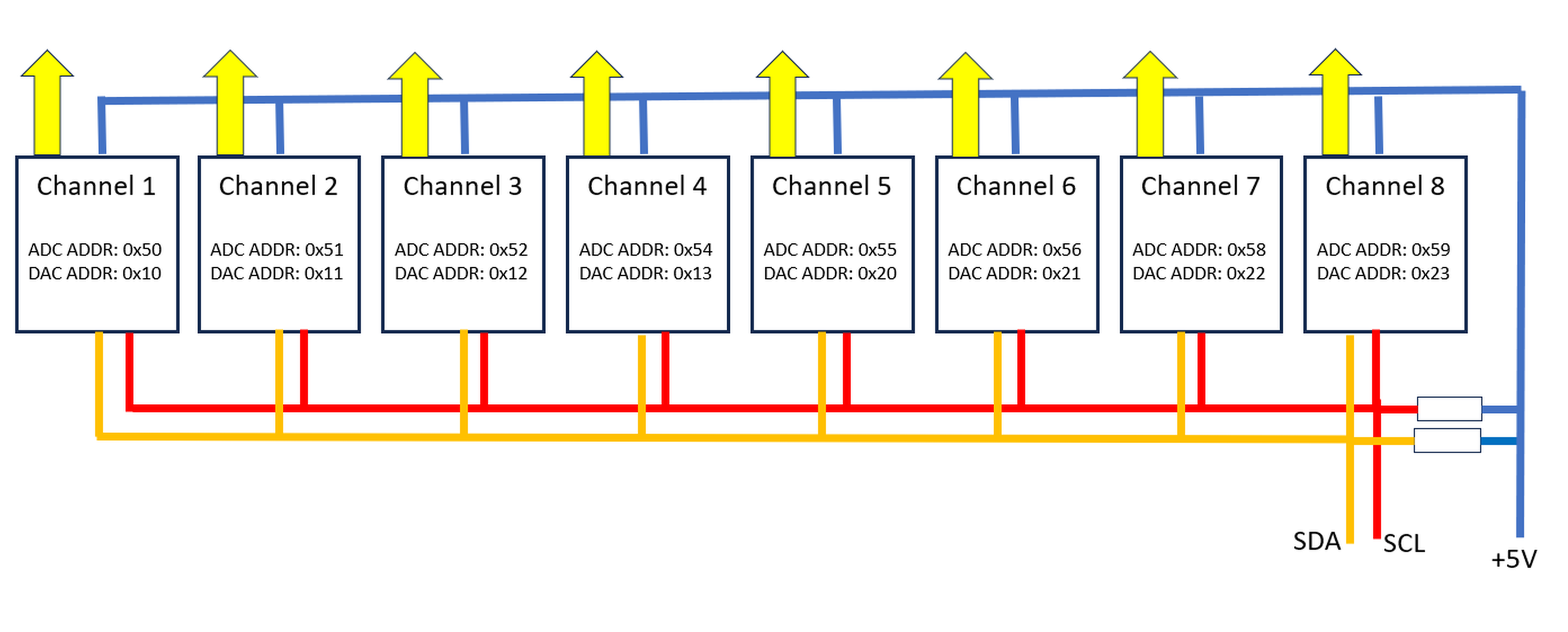} 
    \captionof{figure}{Block diagram illustrating the I2C-based control architecture for PMT high-voltage adjustment. } 
    \label{fig:jumper_control} 
\end{figure}

Selected PMT modules have built-in high-voltage (HV) generators and require only +5V to operate. In addition, the HV level can be adjusted via a control voltage pin. To achieve digital control of PMT HV levels, 8 independent ADC (ADC121C021, Texas Instruments) and DAC (LTC2631, Analog Devices) loops are used. These loops can individually generate 0--1.8~V output voltage range. The ADC read and DAC output assignments are achieved via Arduino Uno board \footnote{\url{https://www.arduino.cc}} with I2C protocol. I2C addresses of ADC and DAC chips are assigned via jumpers as shown in \autoref{fig:jumper_control} and \autoref{tab:jumper_control}. For user friendly control, a serial communication-based voltage query and control algorithm is developed.

\begin{table}[]
\centering
    \resizebox{1\linewidth}{!}{
    \begin{tabular}{|c|c|c|c|c|c|c|c|c|}
    \hline
    \textbf{}    & \textbf{1}     & \textbf{2}     & \textbf{3}     & \textbf{4}     & \textbf{5}     & \textbf{6}     & \textbf{7}     & \textbf{8}     \\ \hline
    \textbf{K4(A0)} & FLOATING & FLOATING & GND      & +Vcc     & FLOATING & GND      & FLOATING & GND      \\ \hline
    \textbf{K2(A1)} & FLOATING & GND      & FLOATING & FLOATING & GND      & +Vcc     & +Vcc     & +Vcc     \\ \hline
    \textbf{K3(C0)} & GND      & GND      & FLOATING & +Vcc     & GND      & +Vcc     & GND      & FLOATING \\ \hline
    \textbf{K1(C1)} & GND      & GND      & GND      & FLOATING & FLOATING & +Vcc     & +Vcc     & +Vcc     \\ \hline
    \end{tabular}
    }
    \captionof{table}{ I2C address assignments for ADC (ADC121C021) and DAC (LTC2631) chips via jumper settings.} 
    \label{tab:jumper_control} 
\end{table} 

\section{Eastern Anatolia Observatory}
\label{sec:loc}
The variation in cosmic ray intensity with altitude has been demonstrated multiple times through studies, contributing to the discovery of cosmic rays \citep{Hess1940}. The cosmic ray flux at sea level is approximately 1--2~particles~cm$^{-2}$~s$^{-1}$, and at 3000 meters it increases by a factor of ten, while at 5000 meters it increases by a factor of twenty \citep{GRIEDER20011,Polatoglu}. The change in cosmic ray intensity, depending on atmospheric density, makes the location of cosmic ray detectors crucial.

\mirya was initially installed at the Istanbul University Observatory Laboratory, located on the Beyaz{\i}t campus, where it was constructed. The coordinates of this location are 41$^{\circ}$00'46.93'' N, 28$^{\circ}$57'49.95'' E, and its altitude is 69~meters. At this altitude, which is quite close to sea level, several initial measurements were conducted over few days. The detector was later relocated to its target site, the Eastern Anatolian Observatory (DAG), where it has been operational since October 19, 2023.

DAG hosts Türkiye’s largest and also the first infrared (IR) telescope with a mirror diameter of 4 meters~\citep{2014SPIE.9145E..47K}. The DAG site is located approximately 35~km from the city center of Erzurum by road, on the Karakaya Hills, which are the western extensions of the Paland\"oken mountain range. The observatory is positioned at 39$^{\circ}$46'43'' N, 41$^{\circ}$13'37'' E at an altitude of 3170 meters \citep{10.1007/s10686-024-09952-w}.  The site's high altitude, low humidity, low turbulence, and favorable atmospheric conditions, make it one of the most important locations in the northern hemisphere for astrophysical studies \citep{Balbay2024}.

The \mirya detector is installed approximately 200~meters south of the DAG telescope, in a basaltic geological formation. Although the DAG telescope, due to its slightly higher position, has partial visibility of the city of Erzurum, the location of \mirya does not have a direct line of sight to the city center. The detector is located in an open highland plain, surrounded by mountainous terrain, with low atmospheric density and very sparse vegetation. The absence of electromagnetic interference at this site significantly enhances the quality of cosmic ray data collection. Moreover, previous measurements conducted at the DAG site using a portable muon detector have confirmed the site's suitability for cosmic ray research \citep{polatouglu2023}.

\mirya is housed in a 20-foot dry container (approximately 6.06~m~$\times$~2.44~m~$\times$~2.59~m), which is insulated against external environmental conditions to minimize temperature fluctuations and environmental noise effects on the measurements. Despite the fact that outside weather conditions can reach anywhere from -27 to +25 $^o$~C \citep{Yuzlukoglu2024b} the HVAC system inside the container is set to fix the container temperature at +15$^o$~C and has so far been very successful in that.

\begin{figure*}[t]
        \centering
        \includegraphics[scale=0.4,angle=270]{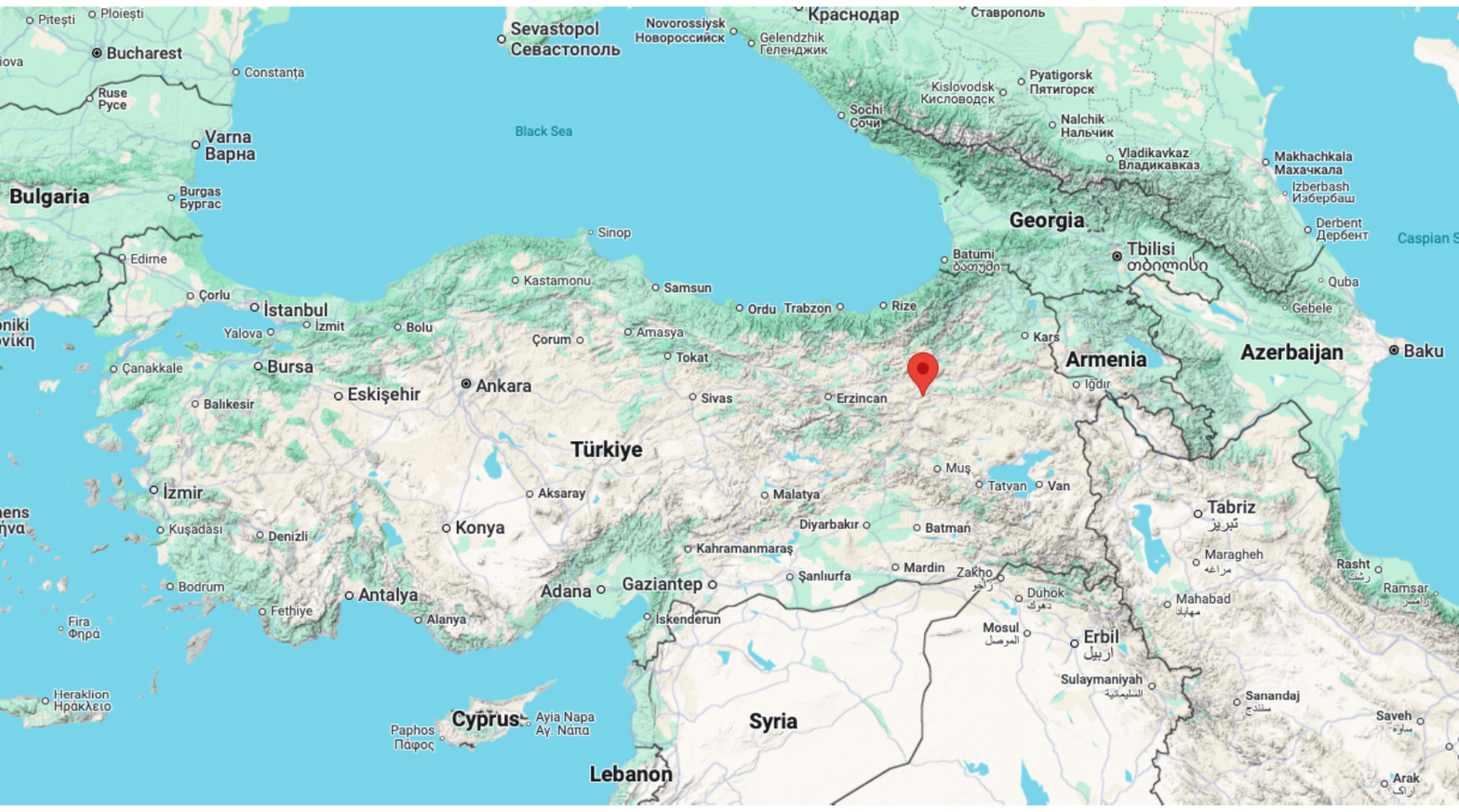} 
        \caption{Location of the Eastern Anatolian Observatory site in Erzurum, T\"urkiye is shown.}
        \label{fig:obsloc}
\end{figure*}

\begin{figure*}[t]
        \centering
        \includegraphics[width=8cm, height=5.5cm]{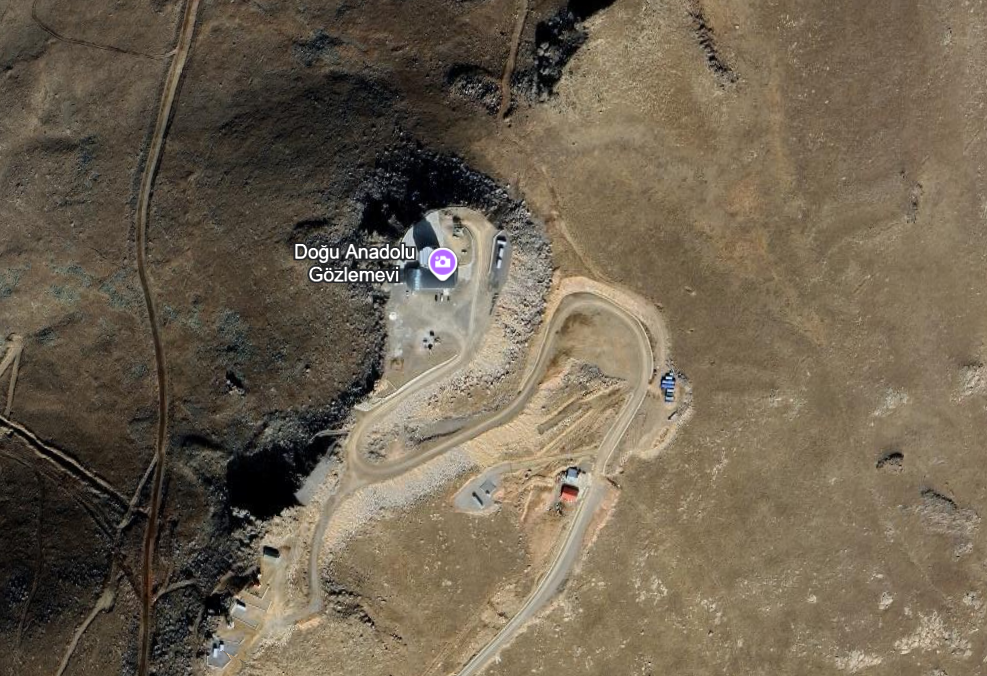} 
        \includegraphics[width=8cm, height=5.5cm]{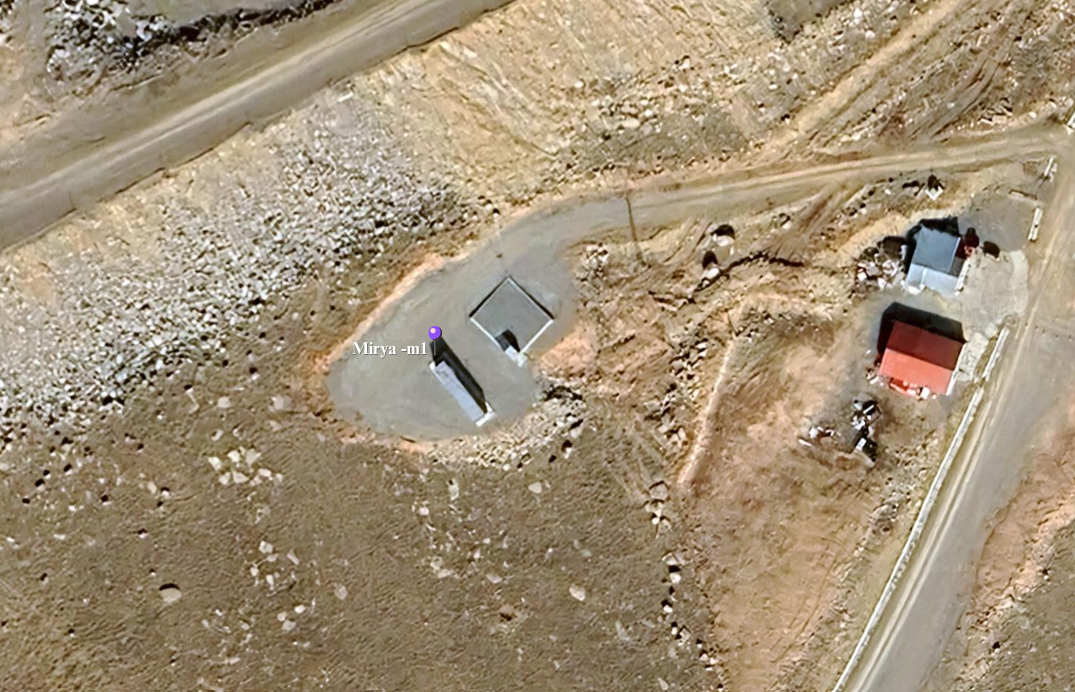}
        \caption{{\it Left Panel:} Eastern Anatolian Observatory site in Erzurum, T\"urkiye. {\it Right Panel:} The container housing the \mirya detector can be seen facing towards 129$^{o}$ from North towards South East.}
        \label{fig:loc}
\end{figure*}

The exact coordinates and altitude of the detector are 39$^{\circ}$46'43'' N, 41$^{\circ}$13'37'' E and 3099.501~meters, respectively (see, \autoref{fig:obsloc}). The location of \mirya at the DAG site and the satellite image of the observatory are shown in \autoref{fig:loc}. Based on the IZMIRAN geomagnetic cutoff rigidity calculator\footnote{\url{https://tools.izmiran.ru/cutoff/}}, the effective cutoff rigidity at the site is $\approx$3.024~GV.

\section{Meteorological Measurements}

 Atmospheric conditions are another factor that must be considered in muon detection. Changes in atmospheric pressure determine whether muons decay before reaching the Earth's surface, or at which altitude they decay. This leads to variations in the rate of cosmic rays detected by muon detectors. Therefore, variations in the number of detected events caused by atmospheric conditions must be corrected \citep{dorman1971, Rigozo2014, Mendon_a_2019}.

 To account for the effects of atmospheric conditions on muon detection, a dedicated weather monitoring station manufactured at our observatory is used at the DAG site inside the container. This station, located at an altitude of 3099 meters, is equipped with high-precision sensors such as the BMP180 and HDC1080, which continuously measure atmospheric pressure, temperature, and humidity. These parameters are recorded synchronously with muon detection events, enabling real-time corrections for atmospheric influences on cosmic ray flux. The recorded weather station data are also made available online to support further analysis. The observatory also has a dedicated site meteorology station, WXT535 by Vaisala, and has been monitoring the atmospheric conditions at the site for a long time \citep{10.1007/s10686-024-09952-w}. We use the pressure measurement by this observatory site to calibrate ours, which is then used to calibrate the cosmic rays as detailed in Subsection \ref{sec:bar_cor}.

\section{First Year Observations}
\label{sec:fyear_obs}

\begin{figure*}[h]
    \centering
    \includegraphics[width=0.8\textwidth]{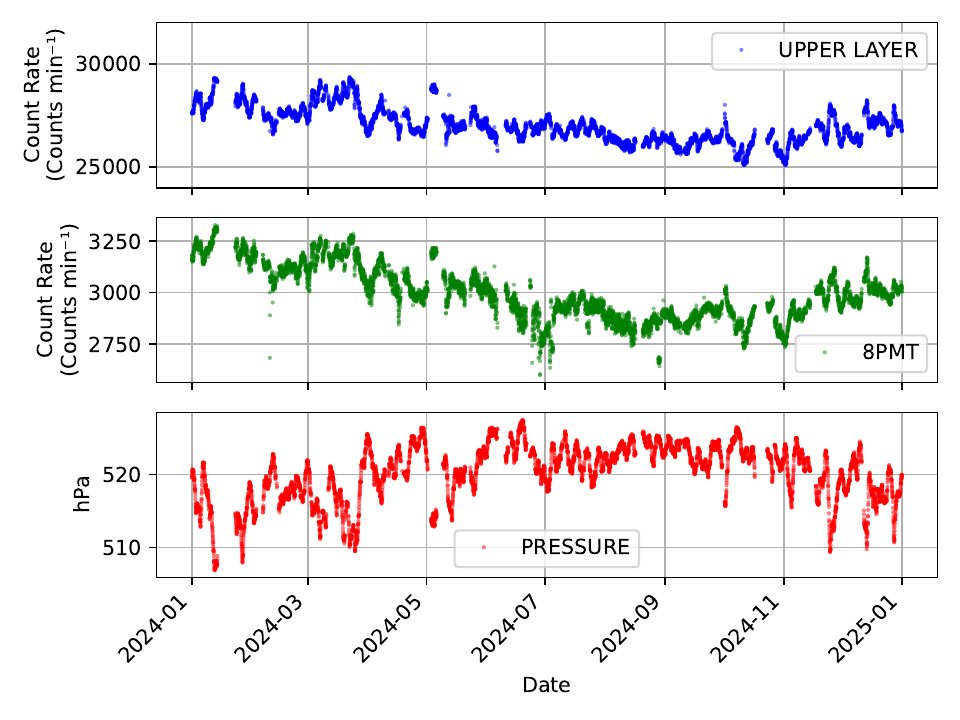}
    \caption{Event rates from the upper layer (upper panel)  when events triggered all the PMTs (middle panel) before pressure correction is applied, and the year long pressure measurements (lower panel). In the plot each point represent the average of sixty measurements obtained every minute.}
    \label{fig:all_data_raw}
\end{figure*}

After two weeks of experimental operations in Istanbul, \mirya has been transferred to its final location at the observatory. Since the 19th of October 2023, it has been operating. We here present the data obtained between January 1st, 2024, up to January 1st, 2025. The raw data obtained when only the PMTs in the upper layer are triggered (relatively lower energy events) and when all the PMTs are triggered (higher energy events) are shown in \autoref{fig:all_data_raw} together with the pressure measurement throughout the year. The detector is operating without any major issues and real-time data can be accessed from a dedicated web page \footnote{\url{http://ist60.istanbul.edu.tr/mirya/}}.

\subsection{Barometric corrections}
\label{sec:bar_cor}
As mentioned above we use two instruments for barometric corrections, one is in the container next to the \mirya and the other one serving for the whole observatory site. Since the barometer in the container worked continuously through the observations presented here, we first calibrated it with the site meteorology station and then used the calibrated inner pressure measurements.

Barometric correction can be applied to cosmic ray data with a relatively simpler approach \citep{ORCA}. The correction is empirically performed using the following equation:

\begin{equation}
    N = N_0 \exp\left(-\beta (P - P_0)\right)
    \label{eq:barometrik}
\end{equation}

\begin{figure}[h]
    \centering
        \includegraphics[scale=0.4]{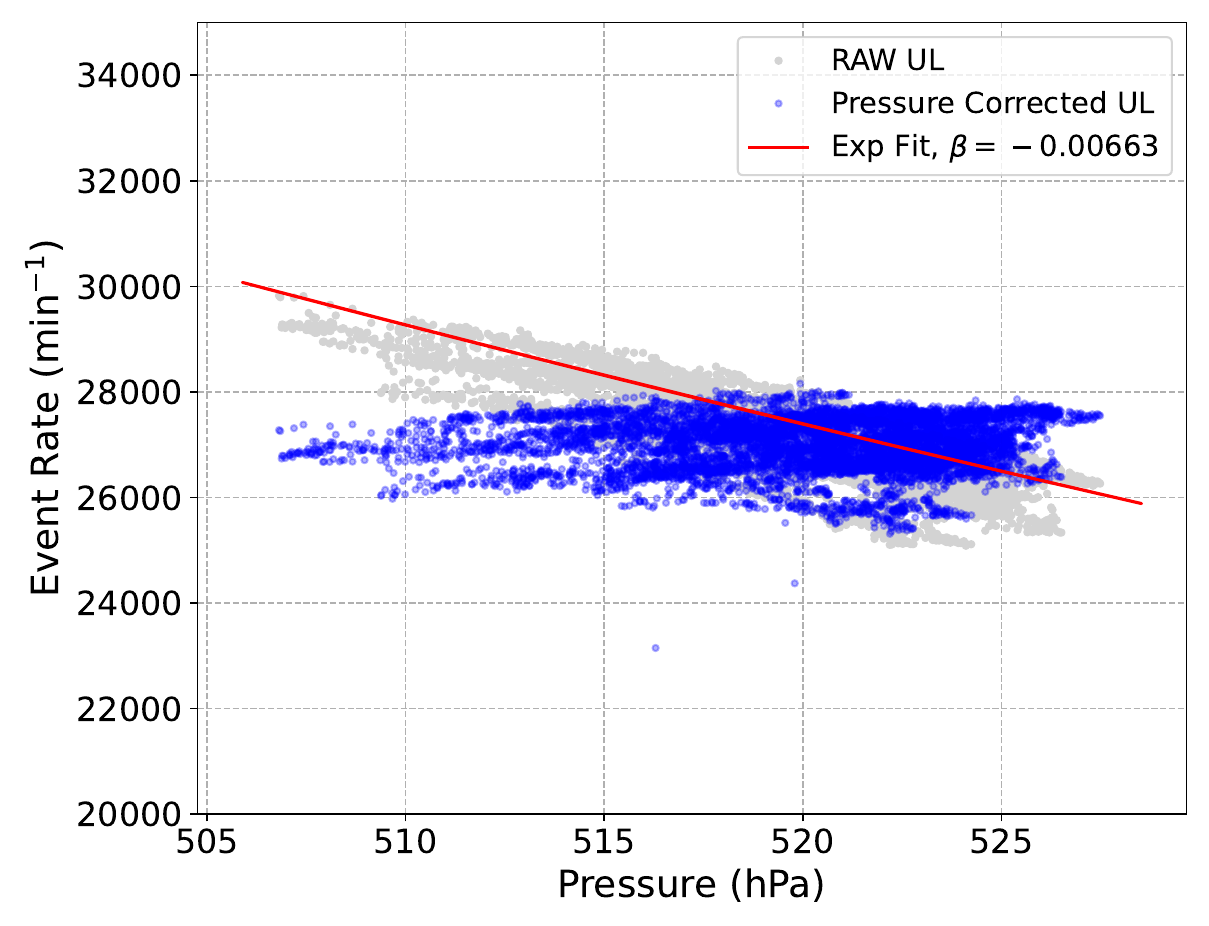}
\includegraphics[scale=0.4]{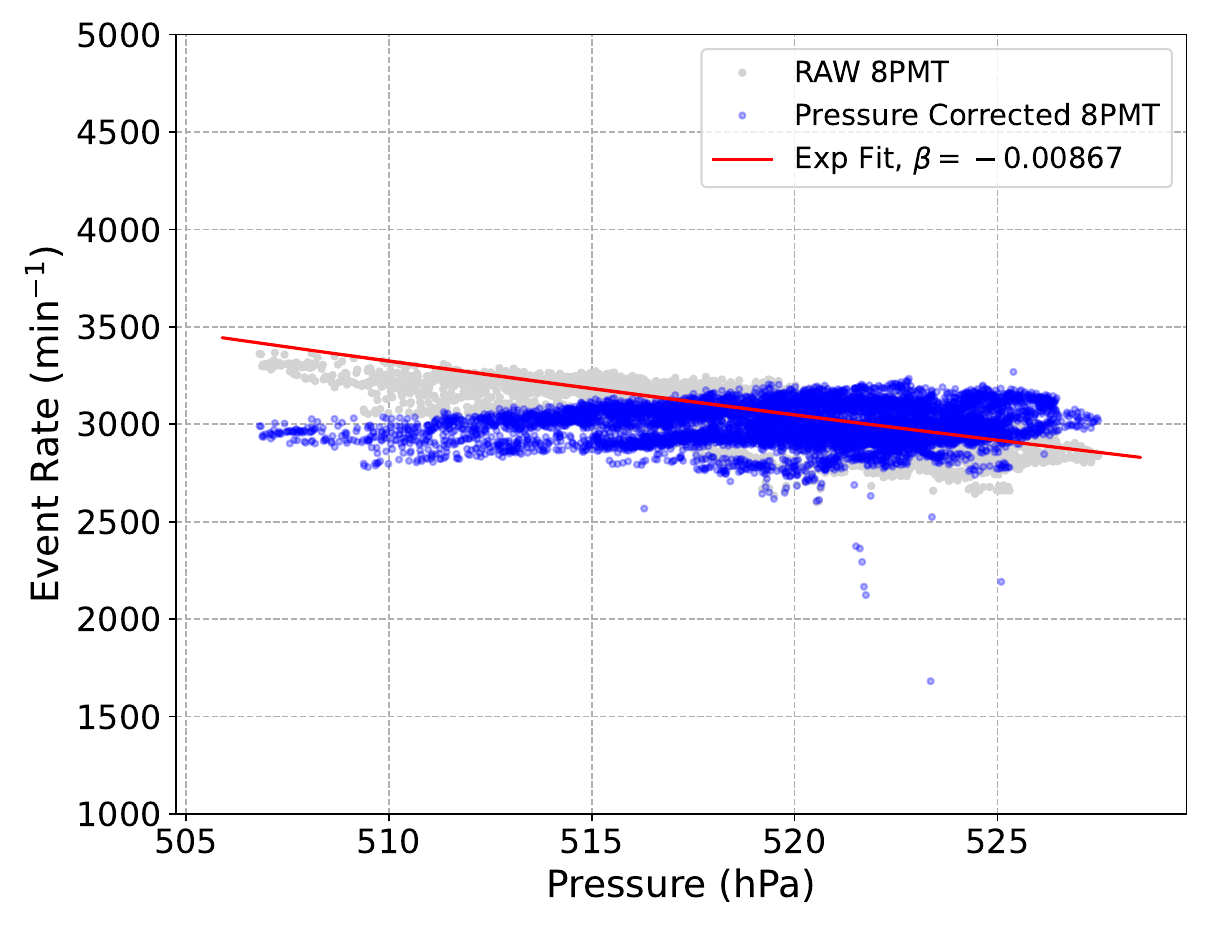}
        \caption{Comparison of pressure correction effects on event rates for the cases when only PMTs in upper layer (upper panel) are triggered and when all the  PMTs are triggered (lower panel). In each plot  
        the grey points represent the raw data, while the blue points show pressure-corrected event rates. 
        The red line indicates the exponential fit with $\beta = -0.00663$ and $\beta = -0.00867$, for upper and lower panel respectively.}
        \label{fig:pressure_correction}
    \end{figure}

\begin{figure*}
    \centering
    \includegraphics[width=0.80\textwidth]{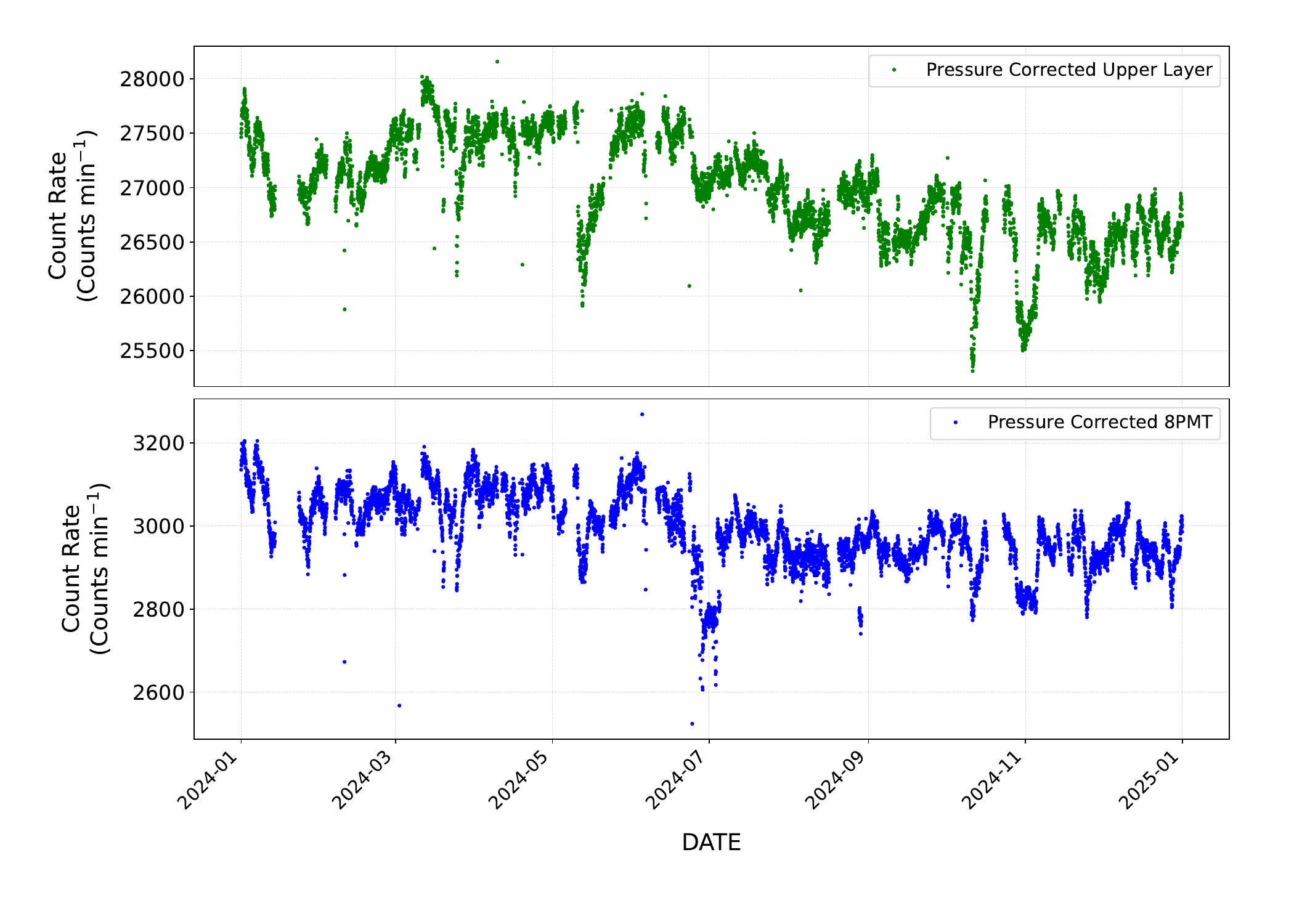}
    \caption{Number of events detected per minute when only the PMTs at the upper layers are triggered (upper panel) and when all of the PMTs are triggered during 2024 after correction for pressure variations is applied. Each point represent the hourly average of measurements obtained every
minute.}
    \label{fig:corrected_full_data}
\end{figure*}

In \autoref{eq:barometrik}, the $\beta$ coefficient is expressed as the barometric coefficient. Here, $N$ represents the corrected count rate, $N_0$ is the uncorrected count rate, $P$ is the pressure value, and $P_0$ is the average pressure value \citep{ORCA}. The value of the $\beta$ coefficient may vary depending on the nature of the measured particle and the altitude of the measurement location \citep{ORCA,dorman1971}. This coefficient can be determined by calculating the slope of the  count rate dependence to the pressure as shown in \autoref{fig:pressure_correction}. The final pressure corrected data is shown in \autoref{fig:corrected_full_data}.

In addition to the effects of atmospheric pressure on the decay altitude of cosmic rays, it has also been shown that atmospheric temperature has significant effects \citep{Duldig}. However, the temperature effect should be considered primarily in long-term measurements. Seasonal temperature variations cause changes in the thickness of the atmosphere, which may result in the measured cosmic ray rate being higher or lower than its actual value. We also checked if the pressure corrected data shows any temperature related dependence, however no significant long term dependence could be found. 

\subsection{Detected Forbush Decreases}

During its first year of operations \mirya already detected a number of Forbush decreases. Two of these events, which coincided with G4 and G5 geomagnetic storms, occurred in March and May 2024. In the following, we summarize these events and provide the general characteristics of the Forbush decreases detected by \mirya. A full discussion of the observations will be presented in a separate study.

\subsubsection{March 2024 Event}

This event involved two active regions (AR 13614 and AR 13615), which produced sympathetic solar flares on March 23, 2024. Among them, an X1.1 class flare occurred from AR3614, peaking at 01:33 UT\footnote{\url{https://solarmonitor.org}}.  A Halo CME with linear velocity of 1470 km/s accompanying the X class flare was detected at 01:25 UT by the LASCO instrument onboard the SOHO. 

The CME impacted Earth's magnetosphere, triggering a G4-level geomagnetic storm on 24 March 2024. On the same day, a significant Forbush decrease was observed  \citep{Mishev,2024Atmos..15.1033M}. It is reported that this Forbush decrease exhibited unusual characteristics, including a significant reduction in galactic cosmic rays and an exceptionally rapid recovery phase \citep{Mishev} . 

\mirya also detected a Forbush decrease during this event on March 24, 2024, count rates dropped by 5.98\% and 7.16\% for the lower-energy and higher-energy events (triggering only the upper layer and all the PMTs, respectively).  It is calculated as the percentage difference between the pre-FD intensity and the minimum intensity observed during the decrease. The recovery time for the 24 March 2024 event was determined to be approximately 1.84 hours for the 8PMT and 1.82 hours for the Upper Layer data, indicating a rapid return to baseline levels. It was also reported to be 1.5 hours at the Oulu neutron monitor station \citep{atmos15091033}. The Forbush Decrease (FD) recovery time here is determined by first identifying the time of minimum intensity and then finding the moment when the intensity reaches 95\% of its average count rate. The recovery time is calculated as shown in Equation~(\ref{eq:recovery_time}):
\begin{equation}
T_{\text{recovery}} = T_{95\%} - T_{\text{min}}
\label{eq:recovery_time}
\end{equation}

\subsection{May 2024 Event}

The G5-level geomagnetic storm of May 2024 was the most intense since the 2003 Halloween storm, marking the strongest event of Solar Cycle 25. This extreme event was triggered by a series of CMEs originating from active region AR 13664, a large and complex sunspot group. This region produced 11 X-class flares, including the largest flare (an X8.7), and at least 10 halo CMEs between May 8 and May 14, 2024 \cite{Hayakawa_2025}.

Three CMEs, associated with X-class flares peaking at 04:37 UT and 21:08 UT on May 8, and 08:45 UT on May 9, were identified as the primary contributors to the extreme geomagnetic storm. After propagating for approximately 42 hours, these CMEs reached Earth, resulting in a severe geomagnetic disturbance from May 10 to 12, 2024. A significant Forbush decrease was also observed during this period \cite{2024EL....14819001C, Hayakawa_2025, 10.5140/JASS.2024.41.3.171, atmos15091033}. 

During the event we detected that the count rates dropped by 8.72\% and 8.28\% for the lower energy and higher energy events (triggering only the upper layer and all the PMTs, respectively). The recovery time for the 10 May 2024 event was determined to be approximately 10.22 hours for the 8PMT and 10.04 hours for the Upper Layer. Here, the calculation for recovery time was performed using Equation \ref{eq:recovery_time}. 

\begin{figure*}
    \centering
        \centering
 \includegraphics[width=0.8\textwidth]{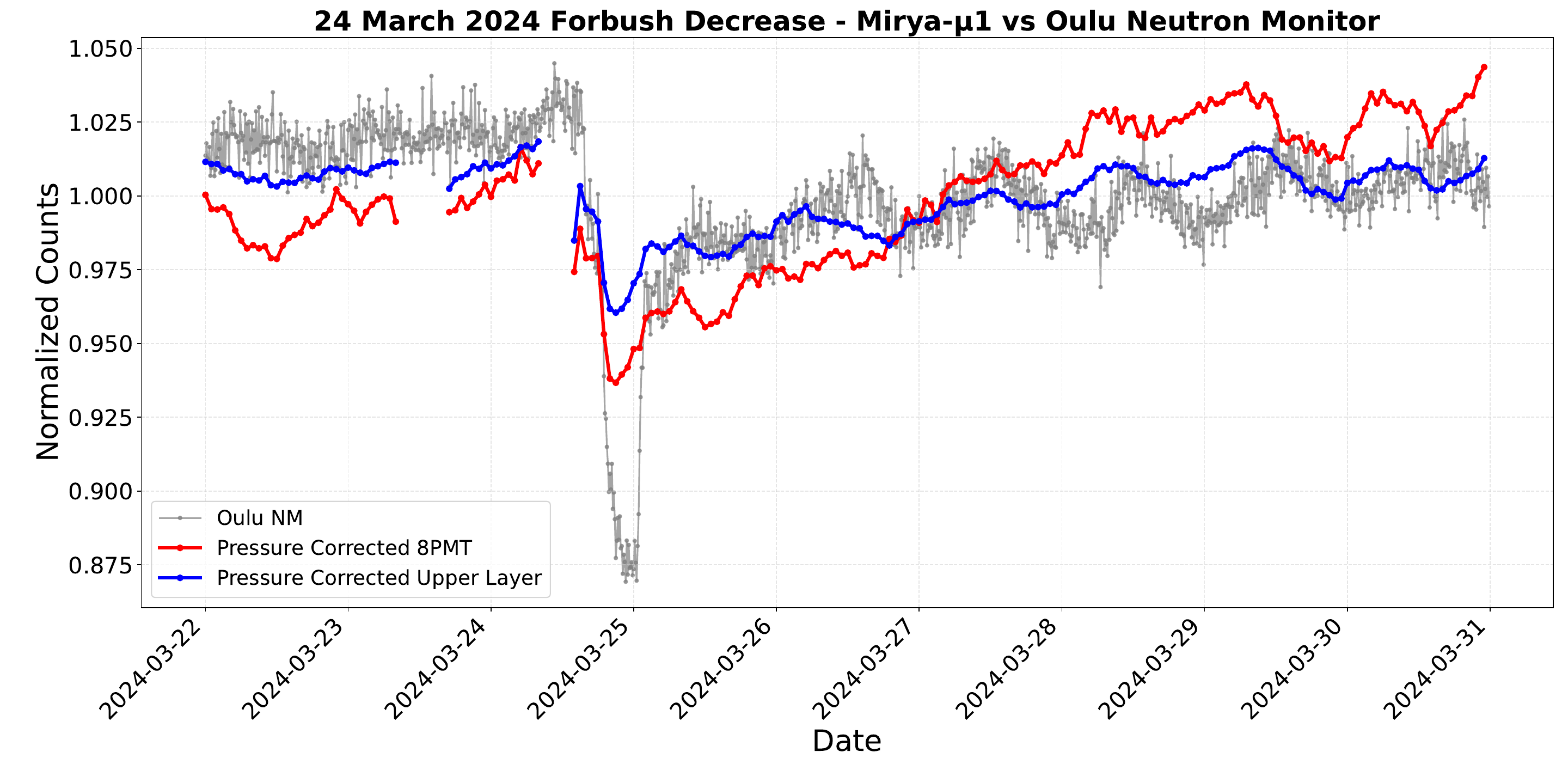}
        \includegraphics[width=0.8\textwidth]{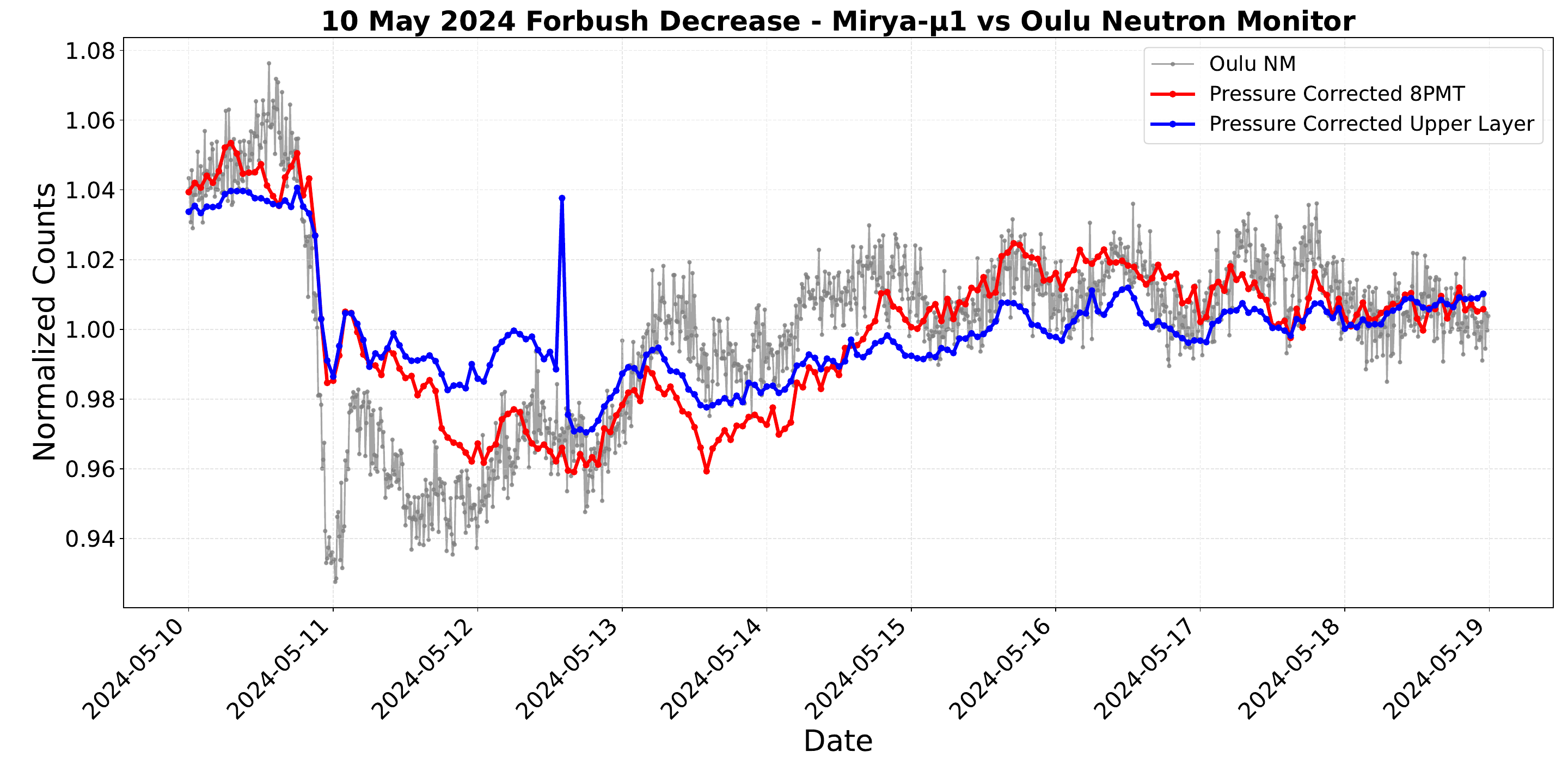}
        \caption{{\it Upper Panel:} 24th March 2024 Forbush decrease observed with \mirya (8PMT and Upper Layer only in red and blue, respectively) and Oulu Neutron Monitor (light grey). The event onset is linked to a high-speed solar wind stream. Unfortunately just hours before the start of the decrease, a power outage on site prevented acquiring data for several hours. {\it Lower Panel:} 10th of May 2024 Forbush decrease observed with \mirya (8PMT and Upper Layer only in red and blue, respectively) and Oulu Neutron Monitor (light grey). The sharp decrease on this date corresponds to a CME-induced Forbush Decrease. The sudden increase seen in the upper layer only data is likely thunderstorm related and caused by low energy particles compared to muons. In both panels data is normalized to the mean rate.}
        \label{fig:forbush}
\end{figure*}

\section{Conclusions}
\label{sec:conc}

In this paper, we introduced a new muon detector, \mirya, located at the T\"urkiye National Observatories' Eastern Anatolia Observatory (DAG) site in Erzurum, T\"urkiye. We present the construction details and components of the detector, emphasizing similarities and differences between \mirya and MITO \cite{MITO}. Having similar detectors in Antarctica and Tenerife may help to compare observations of similar events. 

We also present the first year data. The timing of the \mirya detector's installation coincided with the solar maximum phase of the 25th solar cycle \citep{Schwabe1844}. This alignment is helpful for measuring variations in cosmic ray flux as affected by solar activity since \mirya aims to detect these variations, serving as a precursor to space weather events. On average \mirya detects 27030 events per minute when only the upper layer PMTs are triggered. Most of these events are likely relatively lower energy events. On average, 2993 events trigger all of the PMTs passing through the lead layers, these are the highest energy events \mirya can detect. 

A full discussion of all the events within all datasets is out of the scope and will be discussed in a forthcoming paper. However, we note that within its first year of operations \mirya was lucky to be able to witness two very strong geomagnetic storm events resulting in two Forbush decreases observed by cosmic ray detectors from around the world \citep{atmos15091033,Mishev,Hayakawa_2025}. \mirya detected both of these events and the lightcurves of these two Forbush decreases are shown in \autoref{fig:forbush}. In both cases our measurements are in agreement with earlier observations of Forbush decreases by other muon detectors \citep{GMDB,KARAPETYAN2024106305, Mishev, atmos15091033}. A detailed discussion of these events together with comparison of results obtained by MITO and ORCA (Observatorio de Rayos C{\'{o}}smicos Ant{\'{a}}rtico, \cite{ORCA}) will be presented elsewhere but showcases the \mirya's potential to be used for space weather related research in the future. 

\section{Future Plans}

Given the installation of the \mirya detector during the solar maximum phase of the 25th cycle, similar events causing variation in cosmic ray flux associated with solar activity are expected to be observed. Moreover, in the long-term operation of the \mirya detector we plan to be able to cover an entire solar cycle to sensitively observe the anti-correlation between the solar cycle and muon detection rates \citep{dag2023mirya}. The resulting data archive will provide insights into both short-term and long-term modulations of cosmic ray flux due to solar activity.

Finally, we note that construction of a neutron detector has just begun, which will be located at the same site as \mirya to complement the muon detection measurements. Our plan is to put the neutron monitor on the layers between the two scintillators shown in \autoref{fig:structure}. This neutron monitor system aims to expand research in this field by measuring cosmic rays of different characteristics and allow for the first time Turkish researchers to obtain their own cosmic ray data for space weather-related studies and contribute to the global network of detectors.

\section{Acknowledgments}
We thank the referees of this manuscript very much whose suggestions and comments improved the text significantly.
We appreciate the generous support from AssemCorp Inc. and especially Mr. AVCI. Without their support, the realization of such a large-scale detector would not have been possible. We thank the T\"urkiye National Observatories staff for their efforts in keeping \mirya up and working especially during the harsh winter conditions. T.G. greatly appreciates the hospitality shown at the Georgia Institute of Technology School of Physics, where part of this work is completed. T.G., M.K.D. has been supported in part by the Royal Society Newton Advanced Fellowship, NAF$\backslash$R2$\backslash$180592. T.G., M.K.D., E.A., B.B., M.T.S. S.O., G.G. are supported in part by the Turkish Republic, Directorate of Presidential Strategy and Budget project, 2016K121370. This study was partly funded by the Scientific Research Projects Coordination Unit of Istanbul University Project number FBA-2024-40554. Co-authors Blanco and Ayuso are supported by the project PID2022-140218NB-I00, funded by the Ministry of Science and Innovation. We acknowledge the NMDB database (www.nmdb.eu), founded under the European Union's FP7 programme (contract no. 213007) for providing data, and the PIs of the neutron monitors at: Oulu (Sodankyla Geophysical Observatory of the University of Oulu, Finland). 

\section{Declaration of generative AI and AI-assisted technologies in the writing process}

During the preparation of this work, the author(s) used ChatGPT for translation purposes and for correcting grammatical errors. After using this tool, the author(s) reviewed and edited the content as needed and take(s) full responsibility for the content of the publication.

%%\clearpage
\bibliographystyle{elsarticle-harv}
\bibliography{tez.bib} 

\end{document}